\documentstyle[12pt]{article}

\newcommand{\sect}[1]{\setcounter{equation}{0}\section{#1}\indent}

\begin{document}

\topmargin 0pt
\oddsidemargin 5mm
\def\bbox{{\,\lower0.9pt\vbox{\hrule \hbox{\vrule height 0.2 cm
\hskip 0.2 cm \vrule height 0.2 cm}\hrule}\,}}
\def\a{\alpha}
\def\b{\beta}
\def\g{\gamma}
\def\G{\Gamma}
\def\d{\delta}
\def\D{\Delta}
\def\e{\epsilon}
\def\h{\hbar}
\def\ve{\varepsilon}
\def\z{\zeta}
\def\t{\theta}
\def\vt{\vartheta}
\def\r{\rho}
\def\vr{\varrho}
\def\k{\kappa}
\def\l{\lambda}
\def\L{\Lambda}
\def\m{\mu}
\def\n{\nu}
\def\o{\omega}
\def\O{\Omega}
\def\s{\sigma}
\def\vs{\varsigma}
\def\S{\Sigma}
\def\vphi{\varphi}
\def\av#1{\langle#1\rangle}
\def\pa{\partial}
\def\na{\nabla}
\def\hg{\hat g}
\def\un{\underline}
\def\ov{\overline}
\def\cF{{{\cal F}_2}}
\def\Hsl{H \hskip-8pt /}
\def\Fsl{F \hskip-6pt /}
\def\cFsl{\cF \hskip-5pt /}
\def\ksl{k \hskip-6pt /}
\def\pasl{\pa \hskip-6pt /}
\def\tr{{\rm tr}}
\def\tcF{{\tilde{{\cal F}_2}}}
\def\tg{{\tilde g}}
\def\shalf{\frac{1}{2}}
\def\nn{\nonumber \\}
\def\w{\wedge}

\def\cmp#1{{\it Comm. Math. Phys.} {\bf #1}}
\def\cqg#1{{\it Class. Quantum Grav.} {\bf #1}}
\def\pl#1{{\it Phys. Lett.} {\bf B#1}}
\def\prl#1{{\it Phys. Rev. Lett.} {\bf #1}}
\def\prd#1{{\it Phys. Rev.} {\bf D#1}}
\def\prr#1{{\it Phys. Rev.} {\bf #1}}
\def\prb#1{{\it Phys. Rev.} {\bf B#1}}
\def\np#1{{\it Nucl. Phys.} {\bf B#1}}
\def\ncim#1{{\it Nuovo Cimento} {\bf #1}}
\def\jmp#1{{\it J. Math. Phys.} {\bf #1}}
\def\aam#1{{\it Adv. Appl. Math.} {\bf #1}}
\def\mpl#1{{\it Mod. Phys. Lett.} {\bf A#1}}
\def\ijmp#1{{\it Int. J. Mod. Phys.} {\bf A#1}}
\def\prep#1{{\it Phys. Rep.} {\bf #1C}}


\begin{titlepage}
\setcounter{page}{0}

\begin{flushright}
COLO-HEP-98/408 \\
hep-th/9806178 \\
June 1998
\end{flushright}

\vspace{5 mm}
\begin{center}
{\large On the Supergravity Gauge theory Correspondence and the  
Matrix Model}
\vspace{10 mm}

{\large S. P. de Alwis\footnote{e-mail:
dealwis@gopika.colorado.edu}}\\
{\em Department of Physics, Box 390,
University of Colorado, Boulder, CO 80309.}\\
\vspace{5 mm}
\end{center}
\vspace{10 mm}

\centerline{{\bf{Abstract}}}
 We review the assumptions and the  logic underlying  the derivation  
of DLCQ
Matrix models.
 In particular we try to clarify what remains valid at  finite   $N$,  
the role
of the non-renormalization theorems and higher order terms in the  
supergravity
expansion.  The relation to Maldacena's conjecture is also discussed.  
In
particular the compactification of the Matrix model  on $T_3$ is  
compared to
the $AdS_5\times S_5$
${\cal N}=4$ super
 Yang-Mills duality, and the different role of the branes in the two  
cases is
pointed out.
\end{titlepage}
\newpage
\renewcommand{\thefootnote}{\arabic{footnote}}
\setcounter{footnote}{0}

\setcounter{equation}{0}
\sect{Introduction}

There appear to be  two conjectures on the relation between gauge  
theory
and gravity. One is the Matrix model \cite{bfss} which was originally  
proposed
as a  microscopic theory whose low-energy limit is 11 dimensional
 supergravity.
The other is the more recent conjecture on the relation between gauge
theory and supergravity \cite{ik},\cite{jm},\cite{gkp},\cite{ew}  
whose
clearest manifestation is in the correspondence between ${\cal N} =4$
$SU(N)$ four dimensional Yang-Mills theory and supergravity (string  
theory?)
on a $AdS_5\times S_5$ background. The Matrix model
can also be compactified and in particular on a three torus, it is  
supposed
to be represented by the same  Yang-Mills theory . One of  the  
purposes of this
investigation is to elucidate the connection between the two  
conjectures
\footnote{Recently there have been two papers by S. Hyun \cite{sh} on  
this
issue.
While there is some overlap between the present work and those papers
 our conclusions are  somewhat different especially with regard to  
the
interpretation of the Matrix model on the three torus and the  
corresponding
AdS picture.}. The other purpose is
to understand why finite $N$ calculations work at least in certain  
cases.

In the next section we will review the arguments given in \cite{as},
\cite{ns} for obtaining the Matrix model. In the course of the  
discussion
we will try to be careful about the logic of these arguments by
distinguishing
between what is actually derived and that which is still conjecture.  
In
particular by expanding on arguments given in \cite{sda} we will try
to explain precisely what the connection to supergravity should be.  
We
will also comment on exactly what is achieved by the recently proven
non-renormalization theorems for the model in relation to the  
connection
between gauge theory and gravity.

In the third section we will discuss the correspondence between the  
higher
order terms in the supergravity expansion and the non-renormalization  
theorem.
We will point out that the latter imposes certain regularities in the
supergravity terms and we will also identify the supergravity terms  
from
which certain non-diagonal terms (in the terminology of \cite{bbpt} )
in the Matrix model expansion arise.
In the third section we will briefly review the recent work
 \cite{jm},\cite{gkp},\cite{ew} on the gauge
theory/gravity connection. In particular we will compare and contrast  
this
with the Matrix model conjecture. The natural place for this  is
clearly the $AdS_5\times S_5$ supergravity/string theory, ${\cal N}=4  
$ four
dimensional  Yang-Mills correspondence. In particular we will argue  
that
although
in the interpretation of this connection given in \cite{ew} the gauge  
theory
is located at the boundary of the space-time, in the Matrix model
the whole space is supposed to be the moduli space of the gauge  
theory.
In fact there is a singularity at the origin which is to be  
interpreted
as a break down of the moduli space approximation and is to be  
replaced by
the non-Abelian quantum dynamics. Alternatively from the supergravity
point of view one may regard the singularity
as being resolved by the branes which are sitting there.

\sect{On the Matrix model}
We begin by summarizing  the arguments of  Seiberg \cite{ns} which  
suggest
a connection to D0 quantum mechanics of the Discrete Light Cone  
Quantization
(DLCQ)
 (i.e. the quantization of the theory compactified on a null circle)  
of
M-theory.

 a) A microscopic Lorentz invariant M-theory should include a  
framework
for calculating scattering amplitudes of the fundamental degrees of  
freedom
(the supergravitons ?). At low energies these amplitudes should
 yield 11 dimensional supergravity.
(This is exactly what happens in string theory. There is a Lorentz
covariant formulation, which yields by general arguments on the  
consistent
coupling of spin two fields, the 10 D supergravity low energy  
effective action.
The
challenge in M theory is to find the analog of this.)

b) Given  a theory satisfying a) its compactification on a null  
circle will
yield scattering
amplitudes which at low energies become those of 11 D supergravity  
compactified
on
a null circle.

c) The theory compactified on a null circle (of radius R) is related  
by an
infinite boost to the theory
compactified on a space-like circle. The study of states in DLCQ M  
theory (with
Planck length $l_P$ and finite values of
light cone energy $P_+$) is most conveniently done in terms of a
$\tilde M$ theory compactified on a space-like circle with vanishing  
radius
$R_s$ and a vanishing Planck length $\tilde l_P$  such that
\begin{equation}\label{seilimit}
{R_s\over\tilde l_P^2}={R\over l_P^2},~~{\tilde R_i\over \tilde l_P}=  
{R_i\over
l_P}
\end{equation}
where the right hand sides are fixed.

d) This limit of $\tilde M$ theory is equivalent to string theory in  
a certain
regime. Namely
one where
\begin{equation}\label{limit}
 l_s\rightarrow 0;~~ g^2_{YM}\equiv{1\over l_m^3}={g_s\over  
l_s^3}={R^3\over
l_P^6}
{}~fixed,~~{ R_i\over  l_s^2}={R_i\over l_ml_P}\equiv U_i~fixed.
\end{equation}
In the above we have introduced the string scale $l_S$ and string
coupling\footnote{
Strictly speaking we should consider these as quantities with tildes  
since they
are related
to $\tilde M$ theory rather than to M theory, but since we are not  
going to
discuss
the space like compactification or the M theory it is not essential  
to make the
distinction.} $g_S$ which are related to the $\tilde M$ quantites by
\begin{equation}\label{}
\tilde l_P= g_s^{1/3} l_s,~ R_s= l_s g_s
\end{equation}
This limit is often referred to as the DKPS limit and we will use  
this name for
it . Note that
the radius of the null circle $R$ has no physical significance and we  
may
conveniently
set $R=l_P$ so that  the length scale set by the gauge theory may be  
identified
with the
Planck length, $l_P=l_m$.

e) String theory in the regime defined in d) is given by D0-brane  
quantum
mechanics;
i.e. $U(N)$ quantum mechanics with 16 supercharges where N is the  
number of
D0-branes and this corresponds to the sector with $P_+=N/R$ in the  
original $M$
theory.

In the above list  a)  is clearly influenced by what happens in  
string theory
and
b) is certainly very plausible. c) on the other hand involves an  
infinite boost
and thus
may be problematic but for the purposes of this paper we will assume  
that it is
meaningful. d) involves a hidden assumption that is normally not made  
explicit.
The relation between M theory and string theory is established only  
at the
level
of the effective actions. What is assumed here is that this relation  
holds also
at the
microscopic level. However this is a standard and plausible  
assumption that we
will
not question here.

The real problem is e). The (perturbative) string action (i.e. the
sigma model action) is not defined in this limit (\ref{limit}). In  
fact all
D-brane actions are also ill-defined
in the limit (since the tensions become infinite) except for  the  
D0-brane
action.
If one took the open string representation of the latter, it becomes  
the
quantum mechanics action
\begin{equation}\label{action}
S_{QM} =-{1\over 4g^2_{YM}}\int_{W_{1}}\tr (D_{\a}X_iD^{\a}X_i
+{1\over 4}[X_i,X_j]^2)+fermion~terms.
\end{equation}
since the higher order terms in $\a '$ disappear. Here the $X_i$ are  
the ten
dimensional
gauge fields which in this case are to be interpreted as operators  
governing
the
position fluctuations of the branes.

However it is the closed string representation of this action that is  
directly
related to
the Kaluza-Klein reduction of the 11 D graviton. (see for example  
\cite{pt}).
This action
for a D0-brane in a background field given by the metric $g$ and RR  
field $C$
is
 \begin{equation}\label{daction}
S_{sugra}=-{1\over gl_s}\int dt e^{-\phi}\sqrt{\det g}+{1\over  
l_s}\int C
\end{equation}
One would then expect a relation of the  
form\begin{equation}\label{gagrav}
\int dX'e^{iS_{QM}[U+X']} = \lim_{DKPS} e^{iS_{sugra}[U]}.
\end{equation}
between these two when $g$ and $C$ are
due to a cluster of  D0 -branes and $S_{sugra}$ is the supergravity
representation
of the probe brane action when it is a distance $U=r/l_s^2$ (in units  
with mass
dimension!)
 from the
cluster and moving with velocity $\dot U=v/l_s^2$. It is precisely  
relations of
this sort that must be established if the gauge theory gravity
connection implied by the arguments of  \cite{bfss}, \cite{as},  
\cite{ns} is to
be proven.
The problem is that the supergravity form of the action is meaningful  
when a
massless
closed string representation is valid i.e. when $r/l_s>1$, whereas  
the DKPS
limit
takes us to $r/l_s = l_s U\rightarrow 0$.

The supergravity solution corresponding to N zero branes is given
 by \cite{hs}
\begin{eqnarray}\label{metten}
ds_{10}^2&=&-H_0^{-1/2}dt^2+H_0^{1/2}dx^idx^i \nn
e^{-\phi}&=&H_0^{-3/4},~~~C_t=H_0^{-1}-1.
\end{eqnarray}
where $H=1+h,~h={Nc_0gl_s^7\over r^7}$ and $c_0$ is a known constant
whose value is irrelevant for our purposes. If we lift this solution  
to
11 dimensions using the standard formulae (see for example \cite{pt})
then we get
\begin{eqnarray}\label{meteleven}
ds_{11}^2&=&-(1-h)dt^2-2hdx^{11}dt+(1+h)dx^{11~2}+dx^{i2} \nn
&=&2d\tau dx^{-}+hdx^{-2}+dx^{i2} \nn
&=&e^{-2\bar\phi /3}\bar{ds}_{10}^2+e^{4\bar\phi /3}(dx^-+\bar  
C_{\tau}d\tau)^2
\end{eqnarray}
 where in the last equation,
\begin{equation}\label{}
\bar{ds}_{10}^2=-h^{-1/2}d\tau^2+h^{1/2}dx^{i~2},~e^{-2\bar\phi  
/3}=h^{-1/2},
{}~\bar C_{\tau}
=h^{-1}.
\end{equation}
In particular the ten dimensional metric above is just the  
(asymptotically)
light like
compactified Aichelburg-Sexl \cite{as} metric which can be rewritten  
as.
\begin{equation}\label{metll}
\bar{ds_{10}}^2=l_s^2(-\bar{h}^{-1/2}d\tau^2+\bar{h}^{1/2}(dU^{2}+U^2d 
\O^2_8)),
\end{equation}
where
 \begin{equation}\label{barh}
\bar h =l_s^4 h={c_0Ng^2_{YM}\over U^7}.
\end{equation}
The argument above was given in essence in \cite{bbpt} and elaborated
on in \cite{kk}.

On the other hand let us consider again the 10 dimensional metric
(\ref{metten}) and take the limit (\ref{limit}).
This limit also leads to the light-like compactified M-theory metric
(\ref{metll}) except that we now have $\tau\rightarrow t$.
Thus we might expect that this fact on the supergravity
side of the D0-brane metric is reproduced by the gauge theory on
 the D0-brane in the same limit.
In other words what we should expect is (\ref{gagrav}).
However  as mentioned earlier the problem is that this limit gives us  
a region
of string theory which  takes us to
substring scales where supergravity is not expected to be valid. Thus  
it
is far from obvious that all graviton scattering amplitudes should be
reproduced by the Matrix model.

Let us now review the argument of \cite{sda} in the light of the
above discussion. The idea is to explain the agreement of the
calculation of \cite{bb}, \cite{bbpt} by using string theory as
the interpolating theory connecting supergravity and gauge theory.
In the above mentioned references the gauge theory effective
action was calculated in a background corresponding to a situation   
in which
one brane is separated from the rest by a distance $r$ and moving
with some velocity $v$. In terms of the variables in the gauge
 theory this means that a variable $U=r/l_s^2$ and
 $\dot U=v/l_s^2$ have acquired
 expectation values. In the limit $l_s\rightarrow 0$ with $U$ fixed,
 since $r\rightarrow 0$, the physical separation of the branes are
 below the string scale and are best described by the gauge theory.
 Using dimensional analysis the perturbative expansion is given by  
\cite{bbpt}
\begin{equation}\label{gauge}
C_{I,L}(N)g^{2L-2}_{YM}{\dot U^I\over U^{3L+2(I-2)}}=
C_{I,L}(N){\dot U^2\over g^2_{YM}}
\left ({g^2_{YM}\dot U^2\over U^{7}}\right )^L
\left ({\dot U\over U^2}\right )^{I-2L-2}.
\end{equation}
Before we go onto discuss the argument further it is important to
stress the meaning of the recently proven non-renormalization  
theorem\cite{pss}
in this context. Firstly it is clear purely from the dimensional  
analysis
that the numerical coefficient of a given $\dot U^I\over U^N$ term
can get a contribution only from the $L=(N-2(I-2))/3$ loop level. In  
particular
this means that $\dot U^4/U^7$ term  only gets a contribution from
one loop and that the $\dot U^6/U^{14}$ from two loops. There is no
question of renormalization of these numerical coefficients and so
the agreement of these with supergravity cannot possibly be affected
by going to strong coupling. {\it Thus the non-renormalization  
theorem is
irrelevant for the purpose of explaining this numerical
 agreement with supergravity}. What it does tell us is that the only  
power
 of $U$ which comes with the $\dot U^4$ term is $U^{-7}$ and that the  
only
 one which comes with $\dot U^{6}$ is $U^{-14}$. The relation of this
 fact to supergravity will be discussed in the next section.
  The numerical agreement
 with supergravity still needs to be explained and this is precisely  
what
 was done in \cite{sda}.

The corresponding open string perturbation expansion is obtained
by replacing the coefficients $C_{I.L}(N)$ by functions  
$C_{I.L}(N,l_sU)$
and it was argued in \cite{sda}  that $C_{I.L}(N,0)=C_{I.L}(N)$
\footnote{This fact is true only for configurations such as the one
being considered with some unbroken supersymmetry, see \cite{sda}.}.
On the other hand  for $l_sU={r\over l_s}$ greater than some
critical value (say 1) the physics can be described by closed string
fields. In this region one typically writes the effective action
in a power series in $l_s^2\cal R$ but one may expect it to be
convergent giving some effective action functional $S[g,\phi, C,  
l_s]$
($C$ stands for the RR field). Now in this closed string formalism a
D0-brane is represented by the action (\ref{daction}).

In the configuration that we are considering the closed string fields
 have the
solutions given in (\ref{metten}) to lowest order in $l_s^2$. Suppose  
now the
solution to the exact effective action $S$ is known. This solution  
when
plugged into (\ref{daction}) will have an expansion of the same form  
as
(\ref{gauge}) but with the coefficients $C_{I.L}(N)$  replaced by
functions $C_{I.L}^{SG}(N,l_sU)$. These functions (since they
are obtained from the exact action functional for closed string  
fields)
would be analytic continuations of the corresponding power series  
obtained
from the $\a '$ expansion. Thus they must be the same as  
$C_{I.L}(N,l_sU)$
in the region $l_sU<1$ and in particular at $l_sU=0$. However it  
turns out
that the {\it exact} value of the so-called diagonal coefficients
$C_{2L+2,L}(N)
=C_{2L+2,L}^{SG}(N,0)$ can be calculated simply from the leading term  
of the
closed string expansion. To see this we first need to plug in the
leading order supergravity solution into (\ref{daction}) and then  
take
the limit $l_s\rightarrow 0$. This gives the (finite!)  
result\footnote{This was
first
observed in \cite{mal}}
\begin{equation}\label{sugra}
-{1\over g^2_{YM}}k^{-1}(\sqrt{1-k\dot{U}^2}-1) ,
\end{equation}
where $k\equiv{cg_{YM}^2N\over U^7}$
with $c$ a known constant. Now the important
 point is that  one expects the DKPS limit of the  full
 $\a'$ expansion to go over into the light like compactification of
 the corresponding low energy M-theory expansion
(this is now a quantum M-theory expansion).
 But purely on dimensional grounds
none of the higher derivative terms in the expansion can contribute
to correcting the numerical coefficients of the ``diagonal terms"
which occur in the expansion of (\ref{sugra}) (see \cite{bbpt} and  
the
discussion in the next
section). Thus the analytically
continued value of the  diagonal functions $C_{2L+2.L}(N,l_sU)$ at
the origin $l_s=0$ are given by the leading supergravity values  
obtained
from (\ref{sugra}). This argument then explains why the supergravity
calculation agrees with the loop expansion calculation in gauge  
theory.

Now the above argument did not actually use large $N$. This is just  
as
 well since the calculations of \cite{bb}, \cite{bbpt} were done
 for $N=2$ but they still agreed with supergravity. {\it The reason  
is
 that regardless of the value of N
 only the leading term in the supergravity expansion contributes to
 the diagonal ($I=2L+2$) terms}. Thus one does not need a suppression  
of
 the higher powers of $R$. However there are other comparisons
 between the gauge theory calculations and supergravity
 which involve at least two scales  where
 finite $N$ calculations disagree with supergravity. The classic case  
is the
calculation of
 Dine and Rajaraman \cite{dr}. In this case the argument used above
  does not apply directly (though there may be a generalization of  
it).
 The reason is that in the above discussion
 we have used the limit (\ref{limit}) of the probe action in a  
background
 solution of supergravity corresponding to a cluster of coincident  
D0-branes
  which can be lifted to eleven dimensions and identified with the
 Aichelburg-Sexl metric (averaged over the light like circle).
 In the more complicated case of \cite{dr}
 (and also the cases considered in \cite{dos},\cite{kt})
 there is no corresponding argument whence one can regard the  
scattering
 of three gravitons to three in terms of the action of one probe.
However if recent work \cite{oy} which contradict \cite{dr} is  
correct,
(see also \cite{ffi},\cite{tv}) there is possibly a more general  
argument than
the one given
above that shows agreement between the finite $N$ Matrix model and
arbitrary supergravity processes in a background with
 one light like compactified circle. On the other hand there are
processes \cite{kt} where the finite $N$ argument is definitely  
violated but
agreement is obtained at large $N$. This does not necessarily mean  
that
only the large $N$ result of the Matrix model is reliable. What it  
does mean
is that both bound state effects and higher order supergravity terms  
must
be taken into account when such comparisons are being made. The  
simple
dimensional arguments that enabled us to conclude that only the  
leading
order supergravity term contributes to the diagonal terms for  
instance
may not be valid. In fact as we shall see in the next section  
agreement
of  even the one loop Matrix model calculation with supergravity for  
the two
graviton
to two graviton case requires taking into account the higher  
derivative terms
in the supergravity side. Thus one should not in general expect  
agreement
with just the contributions from the Einstein term.
\sect{On the non-renormalization theorem and supergravity}
In order to get some perspective on this issue\footnote{I would like
to acknowledge the collaboration of E. Keski-Vakkuri
and P. Kraus in this section.}
 it is necessary to recall
some history. In the BFSS paper it was stated after their observation
(based on the calculation of \cite{dkps}) that the $v^4/r^7$ term
\footnote{For convenience in comparing with standard results in the
literature we have reverted back to the standard notation where
$\dot U\rightarrow v,~ U\rightarrow r$.}
in the Matrix model agreed with the 11D supergravity calculation of
two graviton scattering at zero momentum transfer, that a  
non-renormalization
theorem was needed in order to protect this agreement. Since
there was no discussion of $R^4$ and higher derivative
terms on the supergravity side the
point they were making presumably was that since on the supergravity  
side
the calculation gave only the  term $v^4/r^7$ at order $v^4$
(i.e. that there are no other powers of $1/r$)  this should be the
only contribution in the Matrix model as well. The situation is
much more complicated however, since first of all the Matrix model
(or string theory) one loop calculation has an infinite number of
non-vanishing terms. Thus even for agreement with the one loop Matrix  
model
calculation one needs on the supergravity side (an infinite number  
of)
higher derivative terms. In fact we may reverse the logic that led to
the above quoted statement from BFSS and ask what restrictions the
non-renormalization theorems have on the supergravity expansion.

As pointed out in \cite{rt}, comparison with type II strings implies  
that
the M-theory  low energy expansion has (very schematically) the  
following form,
\begin{equation}\label{R}
S\sim \sum^{\infty}_{r=0}l_p^{3r-9}\int ``R"^{3r+1}.
\end{equation}
The inverted commas are a reminder of the fact that in general there  
may
be covariant derivatives as well as Riemann tensors so that the  
counting
is in powers of squared derivatives.
The first term here is the Einstein term. The second term is the by  
now
well-known $R^4$ derivative term \footnote{See \cite{rt} for the  
original
references to this.},
\begin{equation}\label{}
t^{\mu_1...\mu_8}t^{\nu_1...\nu_8}R_{\mu_1..\nu_2}\ldots  
R_{\mu_7..\nu_8}.
\end{equation}
Where $t$ is a rank eight tensor constructed out of the metric.
It is important to note that at the eight derivative level there are
no covariant derivative terms in the action.

First let us note that the structure of this series is exactly what  
is
required for agreement with the Matrix model expansion\footnote{This  
seems to
have been
first observed in \cite{bgl}.}. This is simply because
the expansion  is  in integer powers of $l_p^3$ and
therefore fits in with the expansion in $g_{YM}\equiv{1\over l_m^3}$
since $l_m$ is to be identified with the Planck length. The  
contribution
of the Einstein term  was discussed above and
it gives exactly the diagonal $I=2L+2$ terms in the Matrix model  
expansion.

The comparison  with the Matrix model, of contributions from this  
$R^4$
term, was made in \cite{kk2}(see also \cite{bb2}) where the  basic  
technique
for going beyond the Einstein term was developed.
Let us first briefly review their method.
Write the metric as
\begin{equation}\label{}
ds^2=(\eta_{\mu\nu}+\D_{\mu\nu})dx^{\mu}dx^{\nu}
\end{equation}
where
\begin{equation}\label{}
\D_{\mu\nu}=h_{--}\d_{\mu}^-\d_{\nu}^--\k f_{\mu\nu}
\end{equation}
The first term on the right hand side is the Aichelberg-Sexl metric  
which is an
exact solution to
the string effective action (\ref{R}) (see \cite{bbpt} and references  
therein).
The second term is
a small perturbation due to the probe. Thus we assume that $f<<1$ so  
that the
metric does
 not change  significantly. Substituting in (\ref{R}) we keep only  
the
quadratric terms.
It is important to note that the linear terms vanish since $f=0$  
gives the
Aichelburg-Sexl
metric whtich is an  exact solution to the quantum corrected  
equations of
motion.
Now  for small enough $f$ we can choose the transverse traceless  
gauge for $f$
so that
in particular $(\mu =+,-,i, \tau =x^{+}/2  $ as in section two)
 only $f_{ij}\ne 0$.  The contribution from  the
 $R^4$ term is of the form  (using the $SO(9)\times SO(1,1)$ symmetry  
of the
configuration
and the fact that $h$ depends only on $r=\sqrt{(x^i)^2}$)
is schematically of the form
$\pa_+^2f\pa_+^2f\pa_{\perp}^2h_{--}\pa_{\perp}^2h_{--}$ where the  
subscript
$\perp$ denotes transverse components. Thus we have the equation of  
motion,
\begin{equation}\label{rfour}
(-\pa_+\pa_--\pa_{\perp}^2+h\pa_+^2)f_{ij}+
b\pa_+^4f_{ij}\pa^2_{\perp}h_{--}\pa^2_{\perp}h_{--}=0
\end{equation}
Writing $f\sim e^{ixp}$ we have, solving iteratively  for the  
Routhian,
\footnote{This is the correct object to compute
in order to compare with the gauge theory calculation as argued in
\cite{bbpt}.}
\begin{equation}\label{}
L'=L-p_-\dot x^- =p_i\dot x^i+p_{\tau}={p_-\over
h}(1-\sqrt{1-h_{--}v_{\perp}^2})+\D L'
\end{equation}
The first term here is the exact solution to the Einstein term alone  
and
corresponds to the
diagonal terms of the Matrix model expansion as discussed in the  
previous
section.
In the case considered here we have from  (\ref{rfour}) the result,
\begin{equation}\label{}
\D L'\sim {p_{\tau}^4\over
p_-}(\pa_{\perp}^2h_{--})^2={N_p^3N_s^2v_{\perp}^8\over R^7  
r^{18}}+\ldots .
\end{equation}
In the last step we've used the formulae $p_{\tau}\sim  
{p_{\perp}^2\over
p_-}\sim p_-v^2$ which
are valid to leading order in $h$ and $p_-={N\over R}$. This term
is not ruled out by the non-renormalization theorem (which  only
restricts the $v^4$ and $v^6$ terms). However its $N$ depends  
disagrees with
the naive perturbative N dependence which must go like $N_pN_s^2$. We  
will
find more such disagreements later and we assume that such  
disagreements
are to be expected since bound state effects will almost certainly  
affect
the N dependence of the perturbation series\footnote{We wish to
thank S. Sethi for discussions on this.}.

It is actually easy to see that these $R^4$ terms will not contribute  
to
renormalizing the $v^4$ or
$v^6$  terms. This is because, as can be seen from (\ref{rfour}),
 in order to maintain the SO(1,1)
invariance the term must have four powers of $\pa_+$ and this leads  
to at least
eight powers of
$v$. It should be stressed that the form (\ref{rfour}) obtains,   
because of the
absence of covariant derivative terms in the $R^4$ term.

At this point one might wonder from whence the infinite number of
non-vanishing one-loop terms on
the gauge theory side namely terms like $v^8/r^{15}$  
etc.\footnote{The
coefficient of the
$v^6/r^{11}$ vanishes in the one loop calculation and this can be  
explained by
the non-
renormalization theorem \cite{pss}} come.  This term clearly does not  
arise
from the $R^4$ term
so it has to come from a $R^7$ term or higher order term. It is easy  
to see
that this term
cannot come from a pure (i.e. with no covariant derivatives) term. In  
fact it
comes from a
14 derivative term of the form $R\na^2 R\na^6R$. This leads to a term  
of the
form
\begin{equation}\label{}
\pa_{\perp}^2 f_{ij}\pa_{\perp}^2\pa_+^2f_{ij}\pa_{\perp}^8h_{--}\sim
{N_p^5N_sv^8\over R^7r^{15}}
\end{equation}

Thus we establish that in order to agree even with the one loop  
Matrix model
result the
$R^7$ expression must have covariant derivative terms (unlike the  
$R^4$ term).
The fact that
such terms must exist starting at the 14 derivative level  means that  
there is
no simple
argument on the supergravity side that would correspond to the Matrix  
model
non-renormalization theorem. To put it another way the  
non-renormalization
theorem
on the Matrix model side implies that on the super gravity side  
certain types
of
terms involving covariant derivatives are not allowed. For instance a   
14
derivative
term of the form $R\na^{10}R$ gives a term
$\pa_+f_{ik}\pa_+f_{jk}\pa^{10}_{\perp}\pa_i\pa_j h_{--}$ and
this would  give a contribution proportional to $v^4/r^{19}$ and  
hence if the
Matrix model supergravity correspondence is valid, must vanish by the
non-renormalization
theorem \cite{pss}. Similarly a term of the form $R\na^8R^2$ gives a
contribution $v^6/r^{17}$ and
must also be absent. In general it appears that all terms of the form
$R\na^{6r-2}R$ and
$R\na^{6r-4}R^2$ must be absent in order to have agreement with the  
Matrix
model
non-renormalization theorem.

\sect{The Matrix model and supergravity on $AdS_5\times S_5$}
Now let us try to generalize the arguments of the first part of  
section 2 to
the case
of Matrix models on torii.

The supergravity solution for an (extremal) Dp-brane is given by
 \begin{equation}\label{}
ds^2 =H_p^{-1/2}(-dt^2+\sum_{i =1}^{p}(dx^{i})^2)+H_p^{1/2}
(dr^2+r^2d\O^2_{8-p}).
\end{equation}
for the metric with the dilaton and the RR field taking the values
\begin{equation}\label{}
e^{-2\phi}=g^{-2}H_p^{p-3\over 2},~~C_{0...p}=(H_p^{-1}-1).
\end{equation}
In the above
\begin{equation}\label{}
H=1+{Ng\bar d_pl_s^{7-p}\over r^{7-p}},
\end{equation}
with $\bar d_p$ a known $p$ dependent constant
and $N$ the number of p-branes and
$g$ is the string coupling.
In the weak coupling limit the Dp-brane is described by
some non-Abelian version of the Born-Infeld action whose exact form  
is
currently
unknown. However one can take the limit \cite{jm}
\begin{equation}\label{mald}
\a '\rightarrow 0, {\rm with}~ g^2_{YM}=(2\pi)^{p-2}g_s(\a  
')^{p-3\over 2}
 {\rm fixed}.\end{equation}
 Note that in this limit the gauge field $A$ on the p-brane as well  
as
 the transverse position operator $U$(the 9-p dimensional scalar  
field
 on the brane which is really the transverse components of the 10  
dimensional
 gauge field) are kept fixed. The effective dimensionless coupling  
constant
 of the gauge theory is $g_{eff}\simeq Ng^2_{YM}U^{p-3}$ and the  
theory
 is strongly coupled in the infra-red for $p< 3$ and is weakly  
coupled in
 the infrared for $p>3$ while at $p=3$ we have ${\cal N}=4$ super  
Yang-Mills
 which is a conformal field theory.

 The same scaling may be done in the supergravity solution and gives
\begin{eqnarray}\label{metric}
{ds^2\over l_s^2}&=&{U^{7-p\over 2}\over g_{YM}\sqrt{d_pN}}
(-dt^2+\sum_{i=1}^p(dx^{i})^2)+{g_{YM}\sqrt{(2\pi)^{p-2}d_pN}\over
U^{7-p\over 2}}dU^2 \nn
&+&g_{YM}\sqrt{(d_pN}U^{p-3\over 2}d\O^2_{8-p}.
\end{eqnarray}
where $d_p=(2\pi)^{p-2}\bar d_p$.
These solutions are supposed to be valid if one can ignore both  
string loop
effects and $\a '$ corrections. As discussed in the second paper of  
\cite{jm}
this is possible if the following conditions are satisfied,
\begin{equation}\label{}
\a'{\cal R}\sim  {1\over  
g_{eff}}<<1,~~e^{\phi}\sim{g_{eff}^{7-p}\over N}<<1,~~
g^2_{eff}\equiv Ng^2_{YM}U^{p-3}
\end{equation}

For the case $p=3$ this metric becomes that of $AdS_5\times S_5$.  
From
such arguments (and the agreements that have been shown to
exist between calculations in black hole physics and gauge theory
such as those in \cite{ik})  Maldacena conjectured that gauge theory
 the large N limit is dual in some sense to supergravity
in the above background.
Also including the $O(1/g^2_{YM}N)$ corrections to the strong  
coupling
expansion
in the gauge theory should
be equivalent to including the string corrections  on the above
 supergravity background, while string loop corrections are governed  
by
$g^2_{YM}$.
Actually in this case it has been argued that there are no string  
correction to
this background.
\cite{kr} so one may  even work with small $g^2_{YM}N$.

 Let us now review the  Matrix model argument for relating gauge  
theory and
 gravity after compactifying on a p-torus.
 One starts with the $p=0$ (D0-brane) case of the earlier discussion
 (see section 2). The limit one takes is the same as (\ref{mald}) for
 $p=0$. As we reviewed in section 2 the theory thus obtained is
then interpreted as a microscopic model of M-theory on a light like
circle.
Now while the limit for $p=0$ is the same as the one taken by  
Maldacena
\cite{jm} eqn(\ref{mald}) the interpretation  in the
other cases is somewhat different. On the one hand the higher  
dimensional
branes in M-theory are supposed to be obtained as condensates of
the D0-branes. Secondly the matrix theory description of M-theory  
compactified
on a p-torus is obtained by T-dualizing the D0-brane theory
\cite{wt},\cite{bfss}. Let us compare the latter procedure with the  
above
discussion of duality.

Under compactification on a p-torus (with radii $r_i$) and  
T-dualization,
\begin{equation}\label{}
r_i\rightarrow \s_i={l_s^2\over r_i};~~~g\rightarrow g^{(p)}=
g\prod_{i=1}^p{l_s\over r_i}={l_s^{3-p}\over l_m^3}\prod\s_i.
\end{equation}
where we have put $g_{YM}^2\equiv 1/l_m^3$.
It is important to observe that the  limit $\l_s\rightarrow 0$
in the compactified Matrix model means
in addition to (\ref{limit}) that we keep the radii of the dual torus
$\s_i$ fixed. (This corresponds to holding $U=r/l_s^2 fixed))$.
Doing this Matrix model rescaling in the supergravity
solutions we get the following:
\begin{equation}\label{}
H_p=1+{Nd_p\prod^p\s_i\over l_s^4l_m^3}{1\over U^{7-p}}\rightarrow
{Nd_p\prod^p\s_i\over l_s^4l_m^3}{1\over X^{7-p}}.
\end{equation}
where $X=l_m^2U$. Rescaling the metric $ds^2\rightarrow {l_m^2\over  
l_s^2}ds^2$
 we have
\begin{equation}\label{mmmetric}
{ds^2}\rightarrow  {X^{7-p \over 2}\over R^{7-p\over 2}}
(-dt^2+\sum_{\a=1}^p(dx^{\a})^2)+{R^{7-p\over 2}\over X^{7-p\over 2}}
(dX^2+X^2d\O^2_{8-p}).
\end{equation}
where
\begin{equation}\label{}
R^{p-7}={l_m^{-3}\over Nd_p\prod\s_i}.
\end{equation}
It is instructive to compare this in the case $p=3$ to the
$AdS_5\times S_5$ case considered in \cite{jm}. For this case the
above becomes (rewriting $X\rightarrow U$ in order to conform to  
notation
that seems to have become standard for AdS spaces),
\begin{equation}\label{mmmet}
{ds}\rightarrow {U^2\over R^2}
(-dt^2+\sum_{i=1}^p(dx^{i})^2)+{R^2\over U^2}
(dU^2+U^2d\O^2_{8-p}).
\end{equation}
This metric is locally the same as the metric (for the case $p=3$) in
(\ref{metric}) but it is not the same globally. The reason is that
in this Matrix model case one has actually divided out by a discrete
symmetry which is a sub group of the (apparent) translation isometry
(under $x^{\a}\rightarrow x^{\a}+a^{\a}$ $\a =0,i$)
of the above metric. However the actual (freely acting) isometry  
group
of $AdS_5$ is $SO(4,2)$. The translation isometry has a fixed point  
at
$U=0$. To see this let us it is only necessary to observe that the
above coordinates of the AdS metric are ill-defined at $U=0$.
The $AdS_{p+1}$
space is defined as the hyperboloid
\begin{equation}\label{ads}
-UV+(X^{\a})^2=-R^2.
\end{equation}
embedded in a $p+2$-dimensional space with metric
 \begin{equation}\label{}
ds^2=-dUdV +(dX^{\a})^2.
\end{equation}
The metric in the  form (\ref{mmmet}) is obtained by eliminating $V$  
and
defining
the coordinates $x^{\a}={X_{\a}R\over U}$. The translation symmetry  
of
the $x^{\a}$ clearly have a fixed point at $U=0$ and hence when  
dividing
by a discrete subgroup of this symmetry  in order to get a 3-torus  
one
gets a singularity at $U=0$. Thus the space-time metric that is  
related
to the Matrix model on $T_3$ is not $AdS_5\times S_5$ which is a  
smooth space
but a space which locally looks like it away from $U=0$, but has a  
singularity
at $U=0$.

However this singularity is just the point at which the moduli space
approximation of the gauge theory breaks down. The singularity must
in fact be replaced by full quantum non-abelian description. In  
contrast
to the situation in the non-orbifolded case here it is unclear  
whether there
is a holographic interpretation.  The holographic interpretation in  
the case of
$AdS_5\times S_5$ comes from the ansatz of \cite{ew} (see also  
\cite{gkp}
for a slightly different interpretation)  according
to which the ${\cal N}=4$ superconformal field theory sits on the  
boundary of
the AdS space
and the correlation functions of the former are obtained from the  
bulk
supergravity
by using the relation (in a Euclidean signature)
\begin{equation}\label{wit}
\int [dA]e^{-S_{CFT}[\phi_0,A]}=e^{-S[\phi ]} .
\end{equation}
 The functional integral is over all gauge theory variables and  
$\phi$ is a
classical fluctuation around the background AdS space which has
boundary value $\phi_0$. The left hand side of this equation is the  
generating
functional
for connected correlation functions and $\phi_0$ is an external  
source which
uniquely
determines the bulk value $\phi$. Thus the theory in the bulk is  
uniquely
determined by
the theory on the boundary giving a holographic picture of bulk  
physics. It
should also
be noted that  since the space has no singularity there is no need to  
have
branes
anywhere in the space.

By contrast in the Matrix model case the equation which replaces  
(\ref{wit}) is
the analog of (\ref{gagrav})
\begin{equation}\label{?}
\int dX'e^{-S_{MM}[U+X']} = \lim_{DKPS} e^{-S_{sugra}[U]}.
\end{equation}
where (in the present case) $S_{MM}$ is the same gauge theory except  
it is now
on
a three torus and the right hand side is the supergravity  
representation of the
probe
D3 brane in the background space given by (\ref{mmmet}) which is  
singular at
the origin.
The latter is effectively to be replaced by the branes (i.e. the  
Matrix model).
Clearly
it is not straightforward to give this a  holographic interpretation.
\sect{Acknowledgements:} I would like to thank Esko Keski-Vakkuri and  
Per Kraus
for collaboration on the material in section 3 and helpful comments  
on the
manuscript, and Nathan Seiberg, Savdeep Sethi and Edward
 Witten for discussions. I would also
like to thank  Edward Witten for hospitality at the Institute for
Advanced Study where much of this work was done and  the
Institute for Theoretical Physics, Santa Barbara for hospitality  
during the
workshop on
Duality where this project originated. Finally I wish to thank  the  
Council on
Research and Creative Work
 of the University of Colorado
for the award of a Faculty Fellowship. This work is partially  
supported by
the Department of Energy contract No. DE-FG02-91-ER-40672.



\begin{thebibliography}{99}
\bibitem{bfss}T. Banks, W. Fischler, S. Shenker, L. Susskind,
Phys. Rev. D55 (1997) 5112; hep-th/9610043.
\bibitem{ik}I. Klebanov, Nucl. Phys. B496 (1997) 231: I. Klebanov and  
S.
Gubser,
 Phys. Lett B413 (1997) 41, hep-th/9708005.
\bibitem{jm}J. Maldacena, ``The Large N limit of Superconformal Field
 Theories And Supergravity", hep-th/9711200.  N. Itzhaki,   
J.Maldacena, J.
Sonnenshein, S. Yankielowicz,  ``Supergravity and The Large N Limit  
of Theories
With Sixteen Supercharges"
hep-th/9802042.
\bibitem{gkp}S. Gubser, I. Klebanov and A. Polyakov,``Gauge Theory  
Correlators
from Non-Critical String Theory", hep-th/9802109.
\bibitem{ew}E. Witten, ``Anti-de Sitter Space And Holography".  
hep-th/9802150.
\bibitem{sh}S. Hyun, ``The Background Geometry of DLCQ Supergravity",
hep-th/9802026;
``Background geometry of DLCQ M theory on a p-torus and holography",
hep-th/9805136
\bibitem{as}A. Sen, ``D0 Branes on $T^n$ and Matrix Theory"  
hep-th/9709220.
\bibitem{ns}N. Seiberg,  Phys. Rev. Lett. 79 (1997) 3577,  
hep-th/9710009.
\bibitem{sda}S. de Alwis, `Phys.Lett. B423 (1998) 59, hep-th/9710219.
\bibitem{bbpt}K. Becker, M. Becker, J. Polchinski, A. Tseytlin,   
Phys.Rev. D56
(1997) 3174,
hep-th/9706072.
\bibitem{dkps}M. Douglas, D. Kabat, Pouliot and S. Shenker,  Nucl.  
Phys. B485
(1997) 85, hep-th/9608024.
\bibitem{hs}G. Horowitz and A. Strominger,  Nucl. Phys. B360, (1991)  
197.
\bibitem{pt}P. Townsend,``Four Lectures on M-theory" hep-th/9612121.
\bibitem{ais}P. Aichelburg and R. Sexl, Gen. Rel. Grav. 2 (1971) 303.
\bibitem{bb}M. Becker and K. Becker,  Nucl.Phys. B506 (1997) 48,
hep-th/9705091.
\bibitem{kk}E. Keski-Vakkuri and P. Kraus, Nucl.Phys. B518 (1998)  
212,
hep-th/9709122.
\bibitem{pss}S. Paban, S. Sethi, M. Stern, ``Constraints From  
Extended
Supersymmetry in Quantum Mechanics", hep-th/9805018, ``Supersymmetry  
and Higher
Derivative Terms in the Effective Action of Yang-Mills Theories",
hep-th/9806028.
\bibitem{mal}J. Maldacena, ``Branes probing black holes",  
hep-th/9709099.
\bibitem{dr}M. Dine and A. Rajaraman, ``Multigraviton Scattering in  
the
Matrix Model", hep-th/9710174.
\bibitem{dos}M. Douglas, H. Ooguri and S. Shenker, Phys.Lett. B402  
(1997) 36,
hep-th/9702203.
\bibitem{kt}D. Kabat and W. Taylor, ``Spherical membranes in Matrix  
theory"
hep-th/9711078;
``Linearized supergravity from Matrix theory", hep-th/9712185.
\bibitem{oy}Y. Okawa and T. Yoneya, ``Multi-Body Interactions of  
D-Particles in
Supergravity and Matrix Theory" hep-th/9806108.
\bibitem{ffi}M. Fabbrichesi, G. Ferretti and R. Iengo, ``Supergravity  
and
matrix theory do not disagree on multi-graviton scattering"  
hep-th/9806018; R.
Echols and J. Gray, ``Comment on multigraviton scattering in the  
Matrix model"
hep-th/9806109; J. McCarthy, L. Susskind,
and A. Wilkins,  ``Large N and the Dine-Rajaraman problem"  
hep-th/9806136.
\bibitem{tv}W.  Taylor IV and M. Van Raamsdonk, ``Three-graviton  
scattering in
Matrix theory revisited", hep-th/9806066.
\bibitem{rt}J. Russo and A. Tseytlin, Nucl.Phys. B508 (1997) 245,
hep-th/9707134
\bibitem{bgl}V. Balasubramanium, R. Gopakumar and F. Larson, ``Gauge  
Theory,
Geometry and the Large N Limit", hep-th/9712077.
\bibitem{kk2}E. Keski-Vakkuri and P. Kraus, ``Short Distance  
Contributions to
Graviton-Graviton Scattering: Matrix Theory versus Supergravity"
hep-th/9712013.
\bibitem{bb2}K. Becker and M. Becker, Phys.Rev. D57 (1998) 6464,  
hep-th/9712238
\bibitem{kr}R. Kallosh and A. Rajaraman, ``Vacua of M-theory and  
string theory"
hep-th/9805041.
\bibitem{wt}W. Taylor, Phys.Lett. B394 (1997) 283, hep-th/9811042.


\end{thebibliography}
\end{document}